\documentclass[aps,prl,reprint,superscriptaddress]{revtex4-1}
\usepackage{graphicx}
\usepackage{amsmath}
\usepackage{bm}
\usepackage{dcolumn}
\usepackage{bm}
\usepackage{amsfonts}
\usepackage{amssymb}
\usepackage[breaklinks=true]{hyperref}
\hypersetup{colorlinks=true, citecolor=blue, filecolor=blue, linkcolor=blue,urlcolor=blue}

\begin{document}
\title{Control of spin current by a magnetic YIG substrate in NiFe/Al nonlocal spin valves}
\author{F. K. Dejene}
\email[e-mail:]{\text{f.k.dejene@gmail.com}}
\affiliation{\textit{Physics of Nanodevices, Zernike Institute for Advanced Materials, University of Groningen, 9747AG, Groningen, The Netherlands}}
\author{N. Vlietstra}
\affiliation{\textit{Physics of Nanodevices, Zernike Institute for Advanced Materials, University of Groningen, 9747AG, Groningen, The Netherlands}}
\author{D. Luc}
\affiliation{\textit{CEA-INAC/UJF Grenoble 1, SPSMS UMR-E 9001, Grenoble F-38054, France}}
\author{X. Waintal}
\affiliation{\textit{CEA-INAC/UJF Grenoble 1, SPSMS UMR-E 9001, Grenoble F-38054, France}}
\author{J. Ben Youssef}
\affiliation{\textit{Université de Bretagne Occidentale, Laboratoire de Magnétisme de Bretagne CNRS,\\ 6 Avenue Le Gorgeu, 29285 Brest, France}}
\author{B. J. van Wees}
\affiliation{\textit{Physics of Nanodevices, Zernike Institute for Advanced Materials, University of Groningen, 9747AG, Groningen, The Netherlands}}
\date{\today}
\begin{abstract}

We study the effect of a magnetic insulator (Yttrium Iron Garnet - YIG ) substrate on  the spin transport properties of Ni$_{80}$Fe$_{20}$/Al nonlocal spin valve (NLSV) devices. The NLSV signal on the YIG substrate is about 2 to 3 times lower than that on a non magnetic SiO$_2$ substrate, indicating that a significant fraction of the spin-current is absorbed at the Al/YIG interface. By measuring the NLSV signal for varying injector-to-detector distance and using a three dimensional spin-transport model that takes spin current absorption at the Al/YIG interface into account we obtain an effective spin-mixing conductance $G_{\uparrow\downarrow}\simeq 5 - 8\times 10^{13}~\Omega^{-1}$m$^{-2}$. We also observe a small but clear modulation of the NLSV signal when rotating the YIG magnetization direction with respect to the fixed spin polarization of the spin accumulation in the Al. Spin relaxation due to thermal magnons or roughness of the YIG surface may be responsible for the observed small modulation of the NLSV signal.
\end{abstract}
\maketitle

The coupled transport of spin, charge and heat in non-magnetic (N) metals deposited on the magnetic insulator Y$_3$Fe$_5$O$_{12}$ (YIG) has led to new spin caloritronic device concepts such as thermally driven spin currents, the generation of spin angular momentum via the spin Seebeck effect (SSE) \cite{uchida_spin_2010}, spin pumping from YIG to metals \cite{castel_platinum_2012}, spin-orbit coupling (SOC) induced magnetoresistance effects \cite{althammer_quantitative_2013,vlietstra_spin-hall_2013} and the spin Peltier effect, i.e., the inverse of the SSE that describes cooling/heating by spin currents \cite{flipse_observation_2014}. In these spin caloritronic phenomena, the spin-mixing conductance $G_{\uparrow\downarrow}$ of the N/YIG interface controls the transfer of spins from the conduction electrons in N to the magnetic excitations (magnons) in the YIG, or \textit{vice versa} \cite{xiao_theory_2010,jia_spin_2011,heinrich_spin_2011,chen_theory_2013,weiler_experimental_2013}. The interconversion of spin current to a voltage employs the (inverse) spin Hall effect in heavy-metals such as Pt or Pd. The possible presence of proximity induced magnetism in these metals is reported to introduce spurious magnetothermoelectric effects \cite{huang_transport_2012,kikkawa_separation_2013} or enhance $G_{\uparrow\downarrow}$ \cite{jia_spin_2011}. Owing to the short spin-diffusion length $\lambda$ in these large SOC metals, the applicability of the diffusive spin-transport model is also questionable. Experimental measurements that alleviate these concerns are however scarce and hence are highly required.


In this article, we investigate the interaction of spin current (in the absence of a charge current) with the YIG magnetization using the NLSV geometry \cite{jedema_electrical_2001,jedema_spin_2003,kimura_temperature_2008}. Using a metal with low SOC and long spin-diffusion length allows to treat our experiment using the diffusive spin-transport model. We find that the NLSV signal on the YIG substrate is two to three times lower than that on the SiO$_2$ substrate, indicating significant spin-current absorption at the Al/YIG interface. By varying the angle between the induced spin accumulation and the YIG magnetization direction we observe a small but clear modulation of the NLSV signal. We also find that modifying the quality of the Al/YIG interface, using different thin-film deposition methods \cite{vlietstra_spin-hall_2013}, influences $G_{\uparrow\downarrow}$ and hence the size of the spin current flowing at the Al/YIG interface. Recently, a low-temperature measurements of a similar effect was reported by Villamor \textit{et al.}\cite{villamor_magnetic_2014} in Co/Cu devices where $G_{\uparrow\downarrow}\sim 10^{11}\Omega^{-1}$m$^{-2}$ was estimated, two orders of magnitude lower than in the literature \cite{heinrich_spin_2011,vlietstra_spin-hall_2013}. Here, we present a room-temperature spin-transport study in transparent Ni$_{80}$Fe$_{20}$ (Py)/Al NLSV devices.

\begin{figure}[t]
	\includegraphics[width=8.5cm]{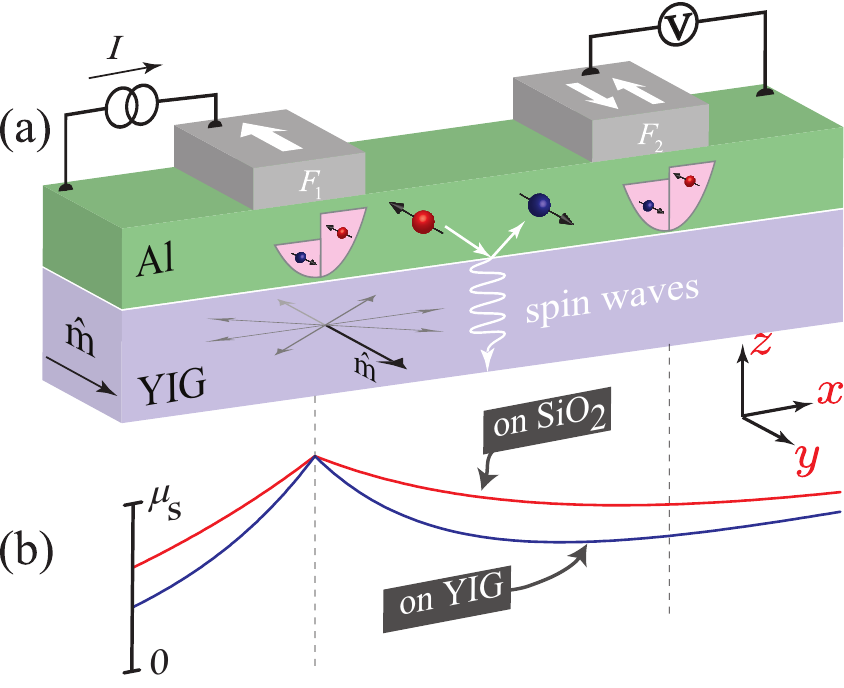}
	\caption{(Color online) Concept of the experiment for $\hat m\parallel\boldsymbol{ \mu_s}$. (a) A charge current through the F$_1$/Al interface creates a spin accumulation $\boldsymbol{\mu}_s$ in the Al. The diffusion of $\boldsymbol{\mu}_s$ to the F$_2$/Al interface is affected by spin-flip relaxation at the Al/YIG interface. Scattering of a spin up electron ($s$=$\hbar/2$) into spin down electron ($s$=$-\hbar/2$) is accompanied by magnon emission ($s$=$\hbar$) creating a spin current that is minimum (maximum) when $\hat{\mu}_s$ is parallel (perpendicular) to the magnetization of the YIG. (b) Profile of $\boldsymbol{\mu}_s$ along the Al strip on a SiO$_2$ (red) and YIG (blue) substrate. The spin accumulation at the F$_2$/Al is lower for the YIG substrate compared to that on SiO$_2$.}
	\label{figure01}
\end{figure}

Figure~\ref{figure01} depicts the concept of our experiment. A nonmagnetic metal (green) deposited on the YIG connects the two in-plane polarized ferromagnetic metals $F_1$ and $F_2$, which are used for injecting and detecting spin currents, respectively. A charge current through the $F_1$/Al interface induces a spin accumulation $\boldsymbol{\mu}_s(\vec{r})=(0,\mu_{s},0)^T$ that is polarized along the $\hat{y}$ direction, parallel to the magnetization direction of $F_1$. This non-equilibrium $\boldsymbol{\mu}_s$, the difference between the electrochemical potentials for spin up and spin down electrons, diffuses to both $+\hat{x}$ and $-\hat{x}$ directions of $F_1$/Al interface with an exponential decay characterized by the spin diffusion length $\lambda_N$. Spins arriving at the detecting $F_2$/Al interface give rise to a nonlocal voltage $V_\text{nl}$ that is a function of the relative magnetic configuration of $F_1$ and $F_2$, being minimum (maximum) when $F_1$ and $F_2$ are parallel (antiparallel) to each other.
\begin{figure*}[tbp]
	\includegraphics[width=18cm]{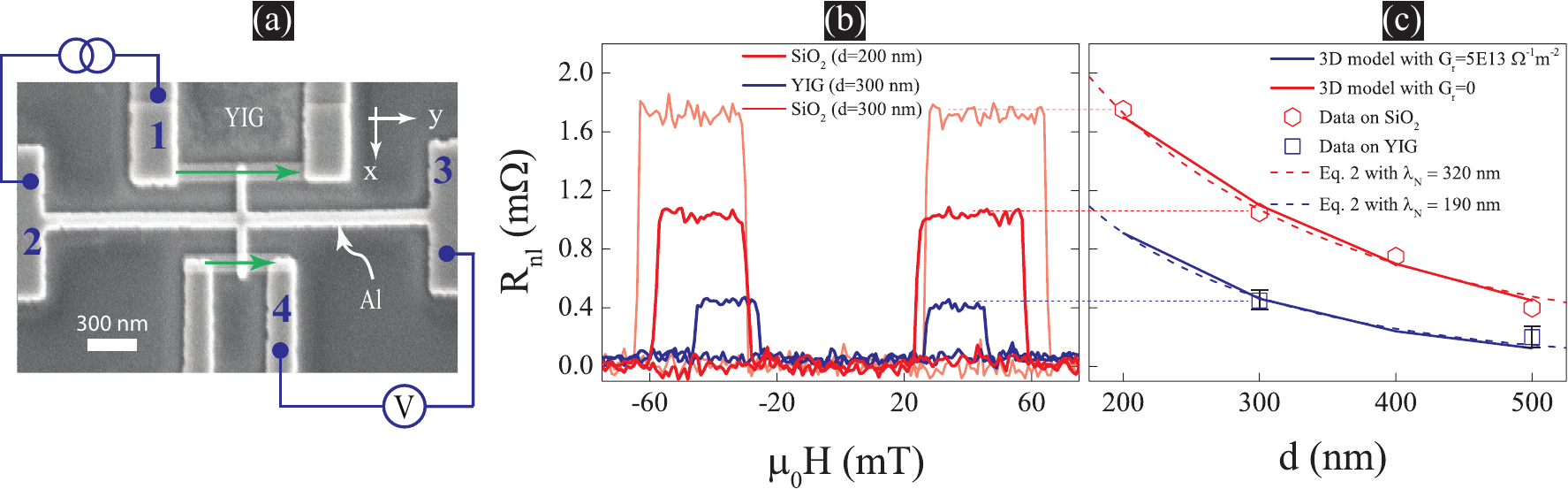}
	\caption{(Color online) (a) Scanning electron microscopy image of the measured Type-A device. Two Py wires (indicated by green arrows) are connected by an Al cross.  A charge current $I$ from contact 1 to 2 creates a spin accumulation at the $F_1$/Al interface that is detected as a nonlocal spin voltage $V_{\text{nl}}$ using contacts 3 and 4. (b) The NLSV resistance $R_{\text{nl}}=V_{\text{nl}}/I$ for representative YIG (blue) and SiO$_2$ (red and orange) NLSV samples. For comparison, a constant background resistance has been subtracted from each measurement. (c) Dependence of the NLSV signal on the spacing $d$ between the injecting and detecting ferromagnetic wires together with calculated spin signal values using a 1D (dashed lines) and 3D (solid lines) spin-transport model. For each distance $d$ between the injector and detector several devices were measured, with the error bars indicating the spread in the measured signal.}
	\label{figure02}
\end{figure*}

For NLSV devices on a SiO$_2$ substrate, spin relaxation proceeds via electron scattering with phonons, impurities or defects present in the spin transport channel, also known as the Elliot-Yafet (EY) mechanism. The situation is different for a NLSV on the magnetic YIG substrate where additional spin relaxation due to thermal magnons in the YIG and/or interfacial spin orbit coupling can be mediated by direct spin-flip scattering or spin-precession. Depending on the magnetization direction $\hat{m}$ of the YIG with respect to $\boldsymbol{\mu}_s$ spins incident at the Al/YIG surface are absorbed ($\hat{m}\perp\boldsymbol{\mu}_s$) or reflected ($\hat{m}\parallel\boldsymbol{\mu}_s$) thereby causing a spin current density $\boldsymbol{j}_s(\vec{r})$ through the Al/YIG interface \cite{chen_theory_2013} 
 \begin{equation}
	\boldsymbol{j}_s(\hat{m})|_{z=0}=G_r\hat{m}\times\left(\hat{m}\times\boldsymbol{\mu}_s\right) +G_i(\hat{m}\times \boldsymbol{\mu}_s)+G_s\boldsymbol{\mu}_s.
	\label{Eq01}
\end{equation}
Here $\hat{m}=(m_x,m_y,0)^T$ is a unit vector parallel to the in-plane magnetization of the YIG, $G_r$ ($G_i$) is the real (imaginary) part of the spin-mixing conductance per unit area and $G_s$ is a spin-sink conductance that can be interpreted as an effective spin-mixing conductance that quantifies spin-absorption (flip) effects that is independent of the angle between $\hat{m}$ and $\boldsymbol{\mu}_s$.

When $\hat{m}\parallel\boldsymbol{\mu}_s$ some of the spins incident on the YIG are reflected back into the Al while some fraction is absorbed by the YIG. The absorption of the spin-current in this collinear case is governed by a spin-sinking effect either due to (i) the thermal excitation of the YIG magnetization (thermal magnons) or (ii) spin-flip processes due to interface spin orbit effects or magnetic impurities present at the interface. This process can be characterized by an effective spin-mixing interface conductance $G_s$ which, at room temperature, is about 20\% of $G_r$ \cite{flipse_observation_2014}. Because of this additional spin-flip scattering, the maximum NLSV signal on the YIG substrate should also be smaller than that on the SiO$_2$. When $\hat{m}\perp\boldsymbol{\mu}_s$ spins arriving at the Al/YIG interface are absorbed. In this case all three terms in Eq.~\eqref{Eq01} contribute to a maximum flow of spin current through the interface. The nonlocal voltage measured at F$_2$ is hence a function of the angle between $\hat{m}$ and $\boldsymbol{\mu}_s$ and should reflect the symmetry of Eq.~\ref{Eq01}. 

Fig.~\ref{figure02}(a) shows the scanning electron microscope image of the studied NLSV device that was prepared on a 200-nm thick single-crystal YIG, having very low coercive field \cite{castel_frequency_2012,castel_platinum_2012,vlietstra_spin-hall_2013}, grown by liquid phase epitaxy on a 500 $\mu$m thick (111) Gd$_3$Ga$_5$O$_{12}$ (GGG) substrate. It consists of two 20-nm thick Ni$_{80}$Fe$_{20}$ (Py) wires connected by a 130-nm thick Al cross. A 5 nm-thick Ti buffer layer was inserted underneath the Py to suppress direct exchange coupling between the Py and YIG. We studied two types of devices, hereafter named Type-A and Type-B devices. In Type-A devices (4 devices), prior to the deposition of the Al (by electron beam evaporation), Ar ion milling of the Py surface was performed to ensure a transparent Py/Al interface. This process, however, introduces unavoidable milling of the YIG surface thereby introducing disordered Al/YIG interface with lower $G_{\uparrow\downarrow}$ \cite{qiu_spin_mixing_2013}. To circumvent this problem, in Type-B devices (2 devices), we first deposit a 20 nm-thick Al strip (by DC sputtering) between the injector and detector Py wires. Sputtering is reported to yield a better interface \cite{vlietstra_spin-hall_2013}. Next, after Ar ion milling of the Py and sputtered-Al surfaces, a 130 nm-thick Al layer was deposited using e-beam evaporation. Similar devices prepared on SiO$_2$ substrate were also investigated. All measurements were performed at room temperature using standard low frequency lock-in measurements.

The NLSV resistance $R_\text{{nl}}=V_\text{{nl}}/I$ as a function of the applied in-plane magnetic field (along $\hat{y}$) is shown in Fig.~\ref{figure02}(b), both for SiO$_2$ (red and orange) and YIG (blue) samples. Note that the magnetizations of the injector, detector and YIG are all collinear and hence no initial transverse spin component is present. The spin valve signal, defined as the difference between the parallel $R_P$ and anti-parallel $R_{AP}$ resistance values, $R_{SV}=R_P-R_{AP}$ on the YIG substrate is about two to three times smaller than that on the SiO$_2$ substrate. This reduction in the NLSV signal indicates the presence of an additional spin-relaxation process even for $\hat{m}\parallel\boldsymbol{\mu}_s$. Assuming an identical spin injection efficiency in both devices, this means that spin relaxation in the Al on the YIG substrate occurs on an effectively shorter spin relaxation length $\lambda_N$. To properly extract $\lambda_{N}$ we performed several measurements for varying distance between the Py wires, as shown in Figure~\ref{figure02}(c) both on SiO$_2$ (red diamond) and YIG (blue square) substrates. Also shown are dashed-line fits using the expression for the nonlocal spin valve signal $R_{SV}$ obtained from a one-dimensional spin transport theory given by \cite{jedema_spin_2003}
\begin{equation}
R_{SV} = \frac{\alpha_F^2 R_N e^{-d/2\lambda_N}}{(\frac{R_F}{R_N}+1)
[\frac{R_F}{R_N}\text{sinh}(d/2\lambda_N)+\text{cosh}(d/2\lambda_N)]}.
\label{eq02}
\end{equation}
Here $R_F=(1-\alpha_F^2)\frac{\lambda_F}{\sigma_F}$ and $R_N=\frac{\lambda_N}{\sigma_N}$ are spin area resistance of the ferromagnetic (F) and non-magnetic (N) metals, respectively. $\lambda_N$ and  $\lambda_F$ are the corresponding spin diffusion lengths, $\sigma_F$ ($\sigma_N$) is the electrical conductivity of the F (N), $\alpha_F$ is the spin polarization of $F$ and $d$ is the distance between the injecting and detecting ferromagnetic electrodes. 
\begin{figure*}[t]
	\includegraphics[width=18cm]{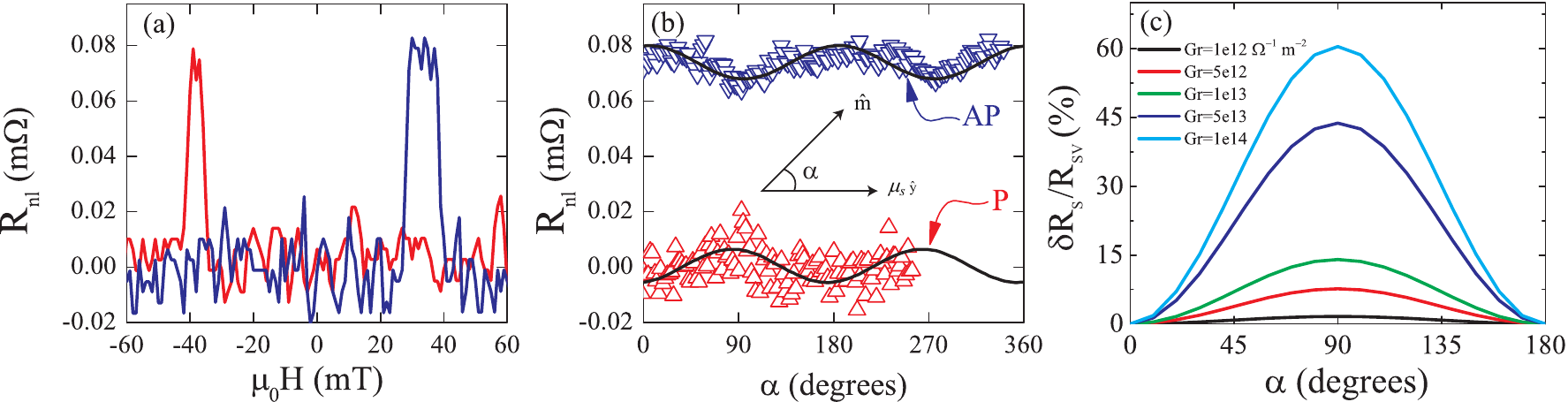}
	\caption{(Color online) (a) Nonlocal spin valve resistance R$_\text{{nl}}$ of a Type-B device with $d$=500 nm between injecting and detecting Py wires and $t_{\text{Al}}$=130 nm. A constant background resistance of 117 m$\Omega$ was subtracted from the original data. (b) Angular dependence of the NLSV signal in the parallel and antiparallel configurations. The AP curve is average of 10 measurements and that of the P state is a single scan. Both resistance states exhibit a $\cos(2\alpha)$ dependence on the angle between $\hat{m}$ and $\boldsymbol{\mu}_s$. The black solid lines are calculated using the 3D-FEM model for $G_r=1\times 10^{13}\Omega^{-1}$m$^{-2}$ that show a percentage modulation of only 12\% corresponding to the green curve in (c) $\delta R_{SV}/R_{SV}$ is plotted. The angular dependent measurement in (b) is from a device for which complete set of measruements were peformed. A spin valves measurement as in (a) was also performed for another device with $d=300$ nm.}
	\label{figure03}
\end{figure*}
Fitting the SiO$_2$ data using Eq.~\eqref{eq02}, we extract $\alpha_F=$0.32 and  $\lambda_{N,\text{SiO}_2}=$320 nm, which are both in good agreement with reported values \cite{jedema_electrical_2001,jedema_spin_2003,kimura_temperature_2008}. A similar fitting procedure for the YIG data, assuming an identical spin injection efficiency, yields an effectively shorter spin-diffusion length $\lambda_{N,\text{YIG}}$=190 nm due to the additional spin-flip scattering at the Al/YIG interface. This value of $\lambda_{N,\text{YIG}}$ therefore contains important information regarding an effective spin-mixing conductance $G_s$ that can be attributed to the interaction of spins with thermal magnons in the YIG. When spin precession, due to the applied external field as well as the effective field due to $G_i$ is disregarded, we can now estimate $G_s$ by relating $\lambda_{N,\text{YIG}}$ to $\lambda_{N,\text{SiO}_2}$ via $G_s$ as (see Supplemental
Material \cite{supplemental}, Sec. I):
\begin{equation}
 \frac{1}{\lambda_{N,\text{YIG}}^2}=\frac{1}{\lambda_{N,\text{SiO}_2}^2}+\frac{1}{\lambda_r^2},
\label{eq03}
\end{equation}
with $\lambda_r^{-2}=2G_s/t_{Al}\sigma_N$ \cite{supplemental}. Using the extracted values from the fit, $\sigma_N$=2$\times 10^{7}$ S/m and $t_{Al}$=130 nm, we extract $G_s\simeq 2.5\times 10^{13}~\Omega^{-1}$ m$^{-2}$, which is about 25\% of the maximum $G_r\sim 10^{14}\Omega^{-1}$m$^{-2}$ reported for Pt/YIG \cite{jia_spin_2011,vlietstra_spin-hall_2013} and Au/YIG \cite{heinrich_spin_2011} interfaces.

To quantify our results we performed three-dimensional finite element simulations using COMSOL Multiphysics (3D-FEM) \cite{slachter_modeling_2011,supplemental} that uses a set of equations that are equivalent to the continuous random matrix theory in 3 dimensions (CRMT3D) \cite{rychkov_crmt3d_2009}. The charge current  $j_c^\alpha
 (\vec{r})$ and spin current $\boldsymbol{j}_s^\alpha(\vec{r})$, (where $\alpha\in{x,y,z}$), are linked to their corresponding driving forces via the electrical conductivity as
\begin{equation}
\begin{pmatrix}j^\alpha_c(\vec{r}) \\ \boldsymbol{j}^\alpha_{s}(\vec{r}) \end{pmatrix} =- \begin{pmatrix} \sigma & \alpha_F\sigma \\ \alpha_F\sigma & \sigma  \end{pmatrix} \begin{pmatrix} \vec{\nabla} \mu_{c} \\ \vec{\nabla} \boldsymbol{\mu}_{s} \end{pmatrix}
\label{eq04}
\end{equation}
where $\mu_c=(\mu_\uparrow+\mu_\downarrow)/2$ and $\boldsymbol{\mu}_s=(\mu_\uparrow-\mu_\downarrow)/2$ are the charge and spin accumulation chemical potentials, respectively. We supplement Eq.~\eqref{eq04} by the conservation laws for charge ($\nabla\cdot j^\alpha_c(\vec{r})=0$) and spin current ($\nabla\cdot\boldsymbol{j}_s=(1-\alpha_F^2)\sigma\left[  \boldsymbol{\mu}_s/\lambda^2+\vec{\omega}_L\times\boldsymbol{\mu}_s\right] $) where $\vec{\omega}_L=g\mu_B\vec{B}/\hbar$ with $g=2$ is the Larmor precession frequency due to spin precession in an in-plane magnetic field $\vec{B}=(B_x,B_y,0)^T$ and $\mu_B$ is the Bohr magneton (see Supplemental
Material \cite{supplemental}, Sec. II). To include spin-mixing at the Al/YIG interface we impose continuity of the spin current $\boldsymbol{j}_s$ at the interface using Eq.~\eqref{Eq01}. The input material parameters such as $\sigma$, $\lambda$ and $\alpha_F$ are taken from Refs.~\onlinecite{bakker_interplay_2010,dejene_spin_2013}.

The calculated spin signals obtained from our 3D-FEM are shown in Fig.~\ref{figure02}(c) for samples on SiO$_2$ (red solid line) and YIG (blue solid line) substrates. By matching the experimentally measured NLSV signal on the SiO$_2$ substrate with the calculated values in the model we obtain $\alpha_F=0.3$ and $\lambda_N=$350 nm. Using these two values and setting $G_s\simeq 5\times 10^{13}~\Omega^{-1}$m$^{-2}$ well reproduces the measured spin signal on the YIG substrate. This value of $G_{s}$ obtained here is consistent with that extracted from our 1D analysis based on Eq.~\ref{eq02}. Hence, the interaction of spins with the YIG magnetization, as modeled here, can capture the concept of spin-mixing conductance being responsible for the observed reduction in the spin signal.

In the following we investigate the dependence of $R_{\text{nl}}$ on the angle $\alpha$ between $\boldsymbol{\mu}_s$ and $\hat{m}$. We rotate the sample under the application of a very low in-plane magnetic field B $\leq$5 mT, enough to saturate the low-coercive ($\leq 0.5$ mT)  YIG magnetization \cite{vlietstra_spin-hall_2013,flipse_observation_2014} but smaller than the coercive fields of $F_1$ and $F_2$ ($\sim$20 mT). This condition is important to maintain fixed polarization axes of $\boldsymbol{\mu}_s$, along the magnetization direction of the injecting ferromagnet, and also have a well defined $\alpha$. The result of such measurement in a Type-B device is shown in Fig.~\ref{figure03}(b) for $d=400$ nm between F$_1$ and F$_2$. Although the measured NLSV signal [Fig.\ref{figure03}(a)] is smaller than in Type-A devices, possibly due to a better Al/YIG interface, $R_{\text{nl}}$ exhibits a $\cos(2\alpha)$ behavior with a maximum (minimum) for $\alpha=0$ ($\alpha=\pi/2$), consistent with Eq.~\eqref{Eq01}. However, the maximum change (modulation) of the signal $\delta R_{s}$=$R_{\text{nl}}(\alpha=0)-R_{\text{nl}}(\alpha=\pi/2))$ is only 12\% of the total spin signal $R_{SV}$, which is at odds with the large spin-mixing conductance estimated from Fig.~\ref{figure02}(b). From anistropic magnetoresistance measurements we exclude the possibility of any rotation of the magnetization of the injector and detector as the cause for the observed modulation in the NLSV signal (see Supplemental
Material \cite{supplemental}, Sec. III-B).

Using the 3D-FEM we calculated the angular dependence of $R_{SV}$ for various values of $G_r$ where the percentage modulation $\delta R_{s}/R_{SV}$ is plotted as a function of $\alpha$, as shown in Fig.~\ref{figure03}(c).  The $G_r$ value of $1\times 10^{13}~\Omega^{-1}\text{m}^{-2}$ extracted from the NLSV signal modulation experiment is one order of magnitude less than reported elsewhere \cite{vlietstra_spin-hall_2013}. This can be possibly caused by the presence of disordered Al/YIG interface with r.m.s. roughness of 0.8 nm (as measured by AFM), which is close to the magnetic coherence volume $\sqrt[3]{V_c}\simeq 1.3$ nm \cite{xiao_theory_2010} of the YIG. This length scale determines the effective width of the Al/YIG interface and also the extent to which spin current from the Al is felt by the YIG magnetization \cite{xiao_theory_2010,schreier_magnon_2013}. Furthermore, the fact that there exists a finite spin-mixing when $\alpha=0$, as discussed above, can also explain the observed small modulation.
\begin{figure}[b]
	\includegraphics[width=8.6cm]{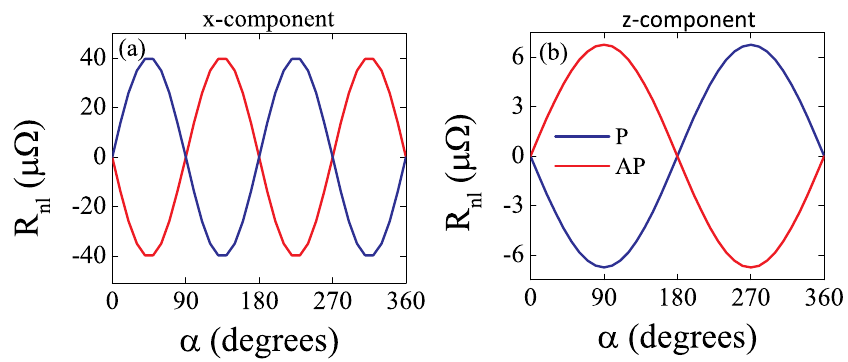}
	\caption{(Color online) Calculated NLSV signals showing the (a) $x$-component and (b) $z$-component of the NLSV signal $R_{nl}$ in the parallel (red) and antiparallel (blue) magnetization configurations of the injector and detector ferromagnetic contacts for $G_r=1\times 10^{13}~\Omega^{-1}$m$^{-2}$ and $G_i=0.1G_r$. Even if the injected spin accumulation is polarized along the magnetization direction of the injecting electrode F$_1$, its interaction with the magnons via the spin-mixing conductance induces these spin accumulation components.}
	\label{figure04}
\end{figure}
It is important to note that in our experiments the non-equilibrium spin accumulation induced by electrical spin injection into Al has a spin-polarization strictly along the direction of the magnetization of $F_1$, which lies along the $\hat{y}$ axis. In the measurement results shown in Figs.~\ref{figure01}(b) and~\ref{figure02}(b) the magnetization of the $F_2$ is always kept either parallel or antiparallel to the detector $F_1$. This ensures that it is only the $\hat{y}$ component of the spin accumulation that is measured in our experiments as it is insensitive to other two spin-polarization directions. It is however possible that the interaction of the initially injected spin accumulation with the YIG magnetization, via $G_{\uparrow\downarrow}$, to induce a finite NLSV signal with components polarized along the $\hat x$- and $\hat z$-directions.

Figure~\ref{figure04} shows the angular dependence of the $\hat x-$ and $\hat z-$ component of the NLSV signal as calculated using our 3D-FEM. While the $\hat{z}$ component exhibits a $\sin(\alpha)$ dependence, the $\hat{x}$ component shows a $\sin(2\alpha)$ dependence which is consistent with Eq.~\eqref{Eq01}. The size of the modulation is determined by $G_r$ for the $\hat x-$ component and by $G_i$ for the $\hat z-$ component. In a collinear measurement configuration these transverse spin accumulation components can induce local magnetization dynamics by exerting a spin transfer torque to the YIG. Separately measuring these spin accumulation using ferromagnetic contacts magnetized along the $\hat{x}$ and $\hat{z}$ directions can be an alternative way to extract $G_{\uparrow\downarrow}$.

In summary, we studied spin injection and relaxation at the Al/YIG interface in Ni$_{80}$Fe$_{20}$/Al lateral spin valves fabricated on YIG. The samples on the YIG substrate yield NLSV signals that are two to three times lower than those grown on standard SiO$_2$ substrates, indicating spin-current absorption by the magnetic YIG substrate. We also observed a small but clear modulation of the measured NLSV signal as a function of the angle between the spin accumulation and magnetization of the YIG. The presence of a disordered Al/YIG interface combined with a spin-flip (sink) process due to thermal magnons or interface spin-orbit effects can be accounted for this small modulation. Using finite element magnetoelectronic circuit theory as well as additional control experiments, we establish the concept of collinear (effective) spin mixing conductance due to the thermal magnons in the YIG. Our result therefore calls for the inclusion of this term in the analysis of spintronic and spin caloritronic phenomena observed in metal/YIG bilayer systems. 
\vspace{1cm}
\begin{acknowledgments}
The authors thank M. de Roosz and J.G. Holstein for technical assistance. This work is part of the research program of the Foundation for Fundamental Research on Matter (FOM) and is supported by NanoLab NL, EU-FET Grant
InSpin 612759 and the Zernike Institute for Advanced Materials.
\end{acknowledgments}
\clearpage
\subsection{\Large  \uppercase{Supplemental Material}}
\subsection{\large {I. Derivation for the effective spin relaxation length in the collinear case}}
\begin{figure*}[t]
	\includegraphics[scale=0.75]{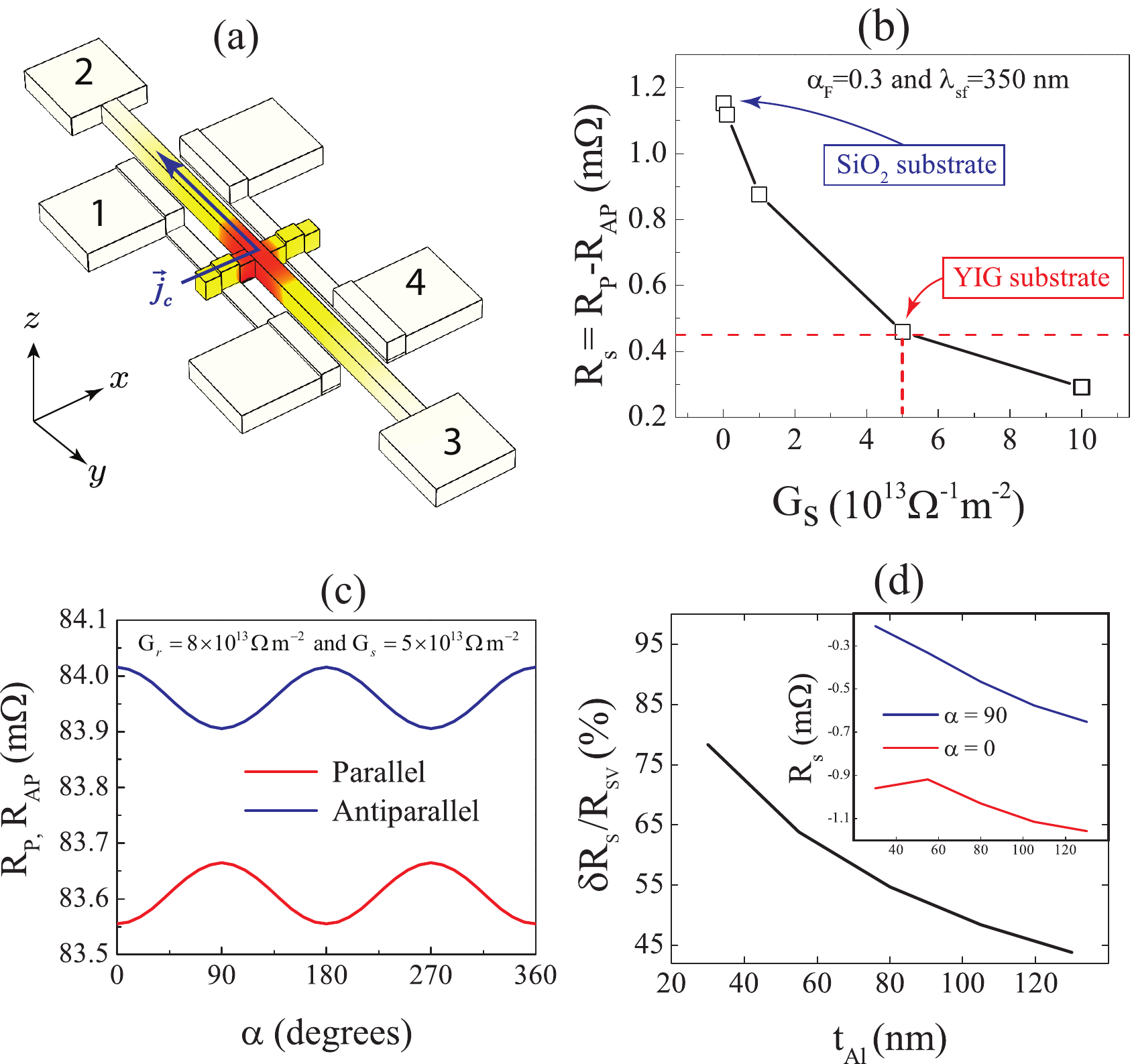}
	\caption{(a) Geometry of the modeled device showing the measurement configuration with a 3D profile and the y-component of the spin accumulation. (b) The dependence of the NLSV signal on the effective (collinear) spin mixing conductance $G_s$. To reproduce the experimentally observed decrease in the spin signal from SiO$_2$ to the YIG substrate, an effective spin mixing conductance of $G_s=5\times 10^{13}~ \Omega^{-1}$m$^{-2}$ is required. (c) The dependence of the NLSV signal on the angle between $\hat{m}$ and $\boldsymbol{\mu}_s$ for $G_s=5\times 10^{13}~ \Omega^{-1}$m$^{-2}$. (d) The dependence of the spin signal modulation amplitude on the thickness of the Al channel signifying the interplay between the spin-mixing conductance and the spin-conductance in the Al channel.}
	\label{figureS1}
	\vspace{-0.5cm}
\end{figure*}

The spin accumulation $\boldsymbol{\mu}_s$, with polarization parallel to the magnetization direction of $F_1$ (see Fig.~S\ref{figureS1}), injected in the Al is governed by the Valet-Fert spin diffusion equation \cite{valet_theory_1993}$
\left[ \partial_x^2+\partial_y^2+\partial_z^2\right] \boldsymbol{\mu}_s=\boldsymbol{\mu}_s/\lambda_N^2$, which can be re-arranged to give
\begin{equation}
	\partial_x^2\boldsymbol{\mu}_s=\boldsymbol{\mu}_s/\lambda_N^2-\partial_z^2\boldsymbol{\mu}_s.
	\label{eqS2}
\end{equation}
Here we assume that, for a homogeneous system, the spin current along the $\hat{y}$-direction is zero. As discussed in the main text, when the YIG magnetization direction $\hat{m}\parallel\boldsymbol{\mu}_s$ the spin current $\boldsymbol{j}_s^{z=0}$ at the Al/YIG interface, in the $\hat{z}$-direction, is governed by the spin sink term $G_s$ in Eq.~1 of the main text. Applying spin current continuity condition at the Al/YIG interface we find that
\begin{equation}
	\frac{\sigma_N}{2}\partial_z\boldsymbol{\mu}_s=G_s\boldsymbol{\mu}_s
	\label{eqS3}
\end{equation}
where $\sigma_N$ is the conductivity of the normal metal. Now after re-arranging Eq.~\eqref{eqS3} to obtain $\partial_z\boldsymbol{\mu}_s$, differentiating it once and using $\partial_z\boldsymbol{\mu}_s=\boldsymbol{\mu}_s/t_{Al}$, where $t_{Al}$ is the thickness of the Al, we obtain
\begin{equation}
	\partial_z^2 \boldsymbol{\mu}_s=-\frac{2 G_s \boldsymbol{\mu}_s}{\sigma_N  t_{Al}}.
	\label{eqS4}
\end{equation}
Substituting Eq.~\eqref{eqS4} into Eq.~\eqref{eqS2} we obtain a modified VF-spin diffusion equation that contains two length scales
\begin{subequations}
	\begin{eqnarray}
		\partial_x^2\boldsymbol{\mu}_s&=&\boldsymbol{\mu}_s/\lambda_N^2 +2 G_r \boldsymbol{\mu}_s/\sigma_N t_{Al}, \label{eqS5a}\\
		&=&\frac{\boldsymbol{\mu}_s}{\lambda_N^2} +\frac{\boldsymbol{\mu}_s}{\lambda_r^2}, \label{eqS5b}   
	\end{eqnarray}
\end{subequations}
where we defined a new length scale $\lambda_r^{-2}=2 G_s/\sigma_N t_{Al}$ that, together with the $\lambda_N$, re-defines an effective spin relaxation length $\lambda_{\text{eff}}^{-2}=\lambda_N^{-2}+\lambda_r^{-2}$. This effective spin relaxation length in the Al channel is weighted by the spin-mixing conductance $G_s$ of the Al/YIG interface. The modulation of the NLSV signal observed in our measurements is hence determined by the interplay between these two length scales, $\lambda_N$ and $\lambda_r$. While the first quantifies the effective spin-conductance of the Al channel ($G_N=\sigma_N A_N/\lambda_{Al}$) over the spin relaxation length, the second is a measure of the quality of the Al/YIG interface and is set by $G_s$. For the devices investigated in this work, using $A_N=t_{Al}w_{Al}$ with the width of the Al channel $w_{Al}=100$nm and $\sigma_{Al}=2\times 10^7$S/m, we obtain $A_N^{-1}G_N\simeq 6\times 10^{13}\Omega^{-1}$m$^{-2}$, which is close to the $G_s$ obtained in our experiments.
This highlights the importance of spin-relaxation induced by the thermal motion of the YIG magnetization, as discussed in the main text. Geometrical enhancement of the modulation can be obtained by reducing $t_{Al}$, as shown in Fig.~S\ref{figureS1}(d),  thereby maximizing spin-absorption at the Al/YIG interface \cite{villamor_magnetic_2014}.  
\vspace{-0.5cm}
\subsection{\large {II. Three dimensional (3D) spin transport model}}
\vspace*{-0.5cm}
Here we describe the our 3D spin transport model used to analyze our data. It is similar to that described in Ref.~\onlinecite{slachter_modeling_2011} for collinear spin transport with the possibility of studying spin-relaxation effects (i) due to the spin-mixing conductance at the Al/YIG interface as well as (ii) Hanle spin-precession due to the in-plane magnetic field [see Sec. III below for detail].
The charge current  $j_c^\alpha
(\vec{r})$ and spin current $\boldsymbol{j}_s^\alpha(\vec{r})$, for $j_c^\alpha(\vec{r})$ (where $\alpha\in{x,y,z}$), are related to the charge $\mu_c(\vec{r})$ and spin potentials $\boldsymbol{\mu}_s$ as 
\begin{equation}
	\begin{pmatrix}j^\alpha_c(\vec{r}) \\ \boldsymbol{j}^\alpha_{s}(\vec{r}) \end{pmatrix} =- \begin{pmatrix} \sigma & \alpha_F\sigma \\ \alpha_F\sigma & \sigma  \end{pmatrix} \begin{pmatrix} \vec{\nabla} \mu_{c} \\ \vec{\nabla} \boldsymbol{\mu}_{s} \end{pmatrix}
	\label{eqS6}
\end{equation}
where $\sigma$ is the bulk conductivity and $\alpha_F$ is the bulk spin polarization of the conductivity. The device geometry we model is shown in Fig.~S\ref{figureS1}(a), showing schematic source-drain configurations as well as voltage contacts. We impose charge flux at contact 1 and drain it at 2. The nonlocal voltage, due to spin diffusion, is obtained by taking the difference between the surface integrated $\mu_c$ at contacts 3 and 4 , both for the parallel (P) or antiparallel (AP) magnetization configurations. To solve Eq.~\eqref{eqS6}, we use conservation laws for charge ($\nabla\cdot j^\alpha_c(\vec{r})=0$) and spin current ($\nabla\cdot\boldsymbol{j}_s=(1-\alpha_F^2)\sigma \boldsymbol{\mu}_s/\lambda^2$) with spin precession due to the in-plane applied field also included in the model. By defining an angle $\alpha$ between $\boldsymbol{\mu}_s$ and the YIG magnetization $\hat{m}$ and allowing for a boundary spin current at the Al/YIG interface using Eq.~1 of the main text, we can study the transport of spins in NLSV devices and their interaction with the YIG magnetization. The material parameters for the model, $\sigma$, $\alpha_F$ and $\lambda_s$ are taken from Ref.~\cite{dejene_spin_2013}. 
\begin{figure*}[tbp]
	\vspace{-0.5cm}
	\includegraphics[scale=0.82]{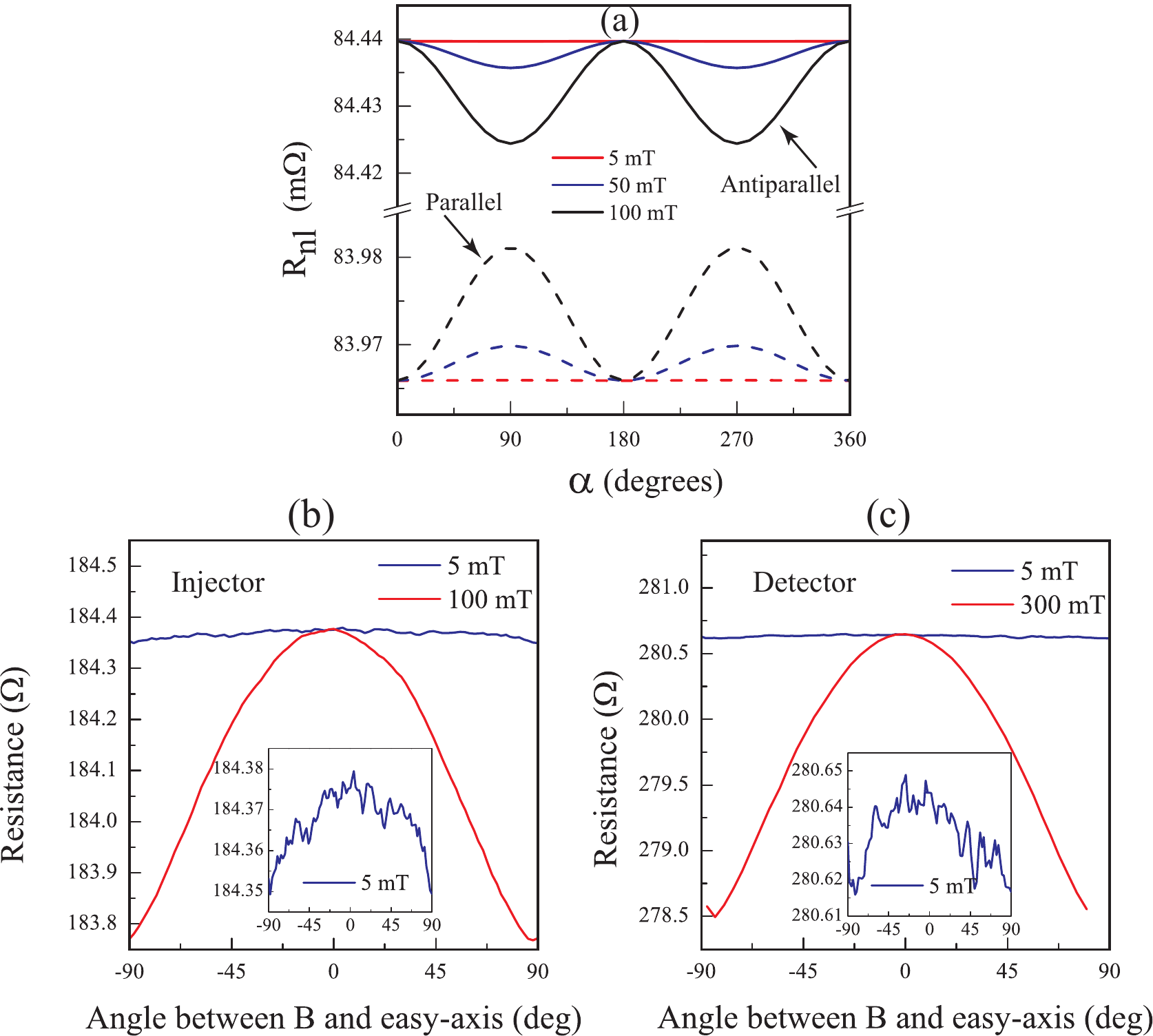}
	\caption{(a) Modulation of the NLSV when only considering the Hanle effect due to the in-plane magnetic field in the P (dashed lines) and AP (solid lines) at 5 mT (red), 50 mT (blue) and 100 mT (black). see text for more details. (b) Anisotropic magnetoresistance (AMR) measurement for the injector (left) and detector (right) ferromagnets at two different magnetic fields. The insets show the full-scale plot of the measurements at 5mT. }
	\label{figureS2}
\end{figure*}
Our modeling procedure involves, first, fitting of the measured NLSV signal on a SiO$_2$ substrate by varying $\alpha_F$ and using $\lambda_N=350$nm. Next, we aim to find $G_s$ of the Al/YIG interface that properly quantifies spin transport properties of the YIG sample. Figure~S\ref{figureS1}(b) shows the dependence of the NLSV signal on $G_s$. As expected, when $G_s$ very low, the NLSV signal is not affected by the presence of the YIG as spins are not lost to the substrate. For $G_s\simeq5\times 10^{13}\Omega^{-1}$m$^{-2}$ we obtain the experimentally measured NLSV signal (shown in red dashed line). For even larger $G_s$ values, the effect is maximum with the NLSV signal falling by almost one order of magnitude. It is important to remember that the value of $G_s$ that is extracted here is a simple measure of spin-flip processes at the Al/YIG interface due to thermal fluctuation of the YIG magnetization or disorder induced effects. At the temperatures of our experiment it is difficult to distinguish which one of the two processes is dominant.

For the angular dependent simulation we only vary the angle $\alpha$ between $\boldsymbol{\mu}_s$ and $\hat{m}$ while keeping all other parameters constant (such as $\alpha_F$, $\lambda_N$ $G_s=5\times 10^{13}~ \Omega^{-1}$m$^{-1}$ and $G_r=8\times 10^{13}~ \Omega^{-1}$m$^{-1}$). As shown in Fig.~S\ref{figureS1}(b) our simulation as described above reproduces the $\cos^2(\alpha)$ dependence observed in our experiments as well as by Villamor \textit{et al.} \cite{villamor_magnetic_2014}. 

For the extracted values of $G_r$ from our analysis, the experimentally observed modulation of the NLSV signal by the rotating magnetization direction of the YIG is small. Possible ways to enhance the modulation are to 1) maximize the spin-mixing conductance via controlled interface engineering of the Al/YIG interface or 2) reduce the thickness of the spin transport channel. In the latter, for a fixed $G_r$, the effect of decreasing the thickness of the spin transport channel is to effectively reduce the spin conductance $G_N$ along the channel thereby maximizing the spin current through the Al/YIG interface. Figure~S\ref{figureS1}(c) shows the thickness dependence of the modulation of the spin signal $\delta R_s=R_s(\alpha=0^0)-R_s(\alpha=90^0)$ normalized by $R_s$ as a function of the thickness $t_{Al}$, with the inset showing that for the P and AP configurations. As the thickness of the Al channel increases the spin current absorption at the Al/YIG interface decreases or vice versa.
\subsection{\large {III. Investigation of possible alternative explanations for the observed modulation}}
It can be argued that the experimentally observed modulation of the NLSV signal can be fully explained by (i) the Hanle spin-precession and/or (ii) the rotation of the magnetizations of the injector/detector electrodes due to the 5 mT in-plane magnetic field. Below, we show that even the combined effect of both mechanisms is too small to explain the experimentally observed modulation of the NLSV signal. 
\subsection{A. Hanle spin-precession induced modulation of the NLSV signal}
Spins precessing around an in-plane magnetic field $\vec{B}$ would acquire an average spin precession angle of $\phi=\omega_L\tau_D$, where $\omega_L=g\mu_B\vec{B}/\hbar$ is the Larmor precession frequency, $\tau_D=L^2/2D_c=25$~ps is the average diffusion time an electron takes to traverse the distance $L$ between the injector and the detector and $D_c=0.005$m$^2$/s is the diffusion coefficient \cite{jedema_electrical_2002}. For an applied field of 5 mT and $L$=500 nm, we obtain $\phi=1.25^o$, giving us a maximum contribution of $1-\cos\phi=\sim$0.02\% [see Eq.~\eqref{eqS7}] to the experimentally observed signal (compared to the $\sim$12\% in Fig.~3(b) of the main text). This is expected because the spin-precession frequency $\omega_L^{-1}$ ($\sim$8 ns) at such magnetic fields is three orders of magnitude slower than $\tau_D$. 

This simple estimate is further supported by our 3D finite element model as we show next. Figure~S\ref{figureS2}(a) shows the angle dependence of the nonlocal signal due to an in-plane magnetic field when we only consider the Hanle effect both for the AP (solid lines) and P (dashed lines) configurations at three different magnetic field values of 5 mT (red), 50 mT (blue) and 100 mT (black). The maximum modulation of the NLSV signal that the Hanle effect presents is only 0.001\% at the measurement field of 5 mT and only become relevant at high fields. Therefore, the Hanle effect alone can not explain the results presented in the main text.
\vspace{-0.5cm}
\subsection{B. Magnetization rotation induced modulation of the NLSV signal}
The in-plane rotation of the sample under an applied magnetic field of 5 mT might induce rotations in the magentization of the injector/detector electrodes. In such a case, a relative angle $\theta_r$ between the magnetization direction of the injector and detector electrodes would result in a modulation of the NLSV signal given by
\begin{equation}
	\frac{\delta R_\text{nl}}{R_\text{nl}(\theta_r=0)}=\frac{R_\text{nl}(\theta_r=0)-R_\text{nl}(\theta_r)}{R_\text{nl}(\theta_r=0)}=\pm|1-\cos\theta_r|,
	\label{eqS7}
\end{equation}
with $+ (-)$ corresponding to the P (AP) configuration. Using Eq.~\eqref{eqS7}, we find that a relative angle $\theta_r\simeq$ 28$^\text{o}$ between the magnetization directions of the injector and detector is required in order to explain the experimentally observed modulation. To determine the field induced in-plane rotation of the magnetization by the applied magnetic field, we carried out angle dependent anisotropic magnetoresistance (AMR) measurements both for the injector and detector electrodes, using a new set of devices with identical dimensions. The AMR measurements were repeated for different magnetic field strengths, at 5 mT and at higher magnetic fields of 100 mT and 300 mT.

Figure S\ref{figureS2}(b) and (c) show the two-probe AMR measurement of the injector and detector electrodes, respectively, at two different magnetic fields. For the injector electrode in Fig.~S\ref{figureS2}(b), at an applied field of 100 mT (red line), an AMR response $\Delta R=R_\parallel-R_\perp=0.6~\Omega$ is observed, where $R_\parallel (R_\perp)$ is the resistance of the ferromagnet when the angle between the applied field and the easy axis is $\theta=0^\text{o}$ $(\theta=90^\text{o})$. For the same electrode, at an applied field of 5 mT (blue line, see also the inset), the AMR response is only 0.025 $\Omega$. Now, by comparing these two measurements we conclude that the effect of the 5 mT field would be to rotate the magnetization of this electrode by a maximum angle $\theta_1=15^\text{o}$ from the easy axis. A similar analysis for the detector electrode, using the AMR responses of 2 $\Omega$ (at 300 mT) and 0.025 $\Omega$ (at 5 mT) in Fig.~S\ref{figureS2}(c), yields a maximum magnetization rotation $\theta_2=10^\text{o}$. Relevant here is the net relative magnetization rotation between the two electrodes $\theta_r=\theta_1-\theta_2=5^\text{o}$ and, using Eq.~\eqref{eqS7}, we conclude that it would only cause a modulation of 0.4 \%, which is much smaller than the 12\% observed in our experiments. Our analysis based on the AMR effect is equivalent to that in Ref.~\onlinecite{villamor_magnetic_2014} where magneto-optical Kerr effect measurements were used to exclude a possible in-plane magnetization rotation as the origin for the observed modulation in the nonlocal spin valve signal \cite{villamor_magnetic_2014}. 

To summarize this section, the Hanle effect and the magnetization rotation induced by the in-plane magnetic field neither separately nor when combined are sufficient to explain the experimentally observed modulation. Only after including the effect of the spin-mixing interaction via $G_{\uparrow\downarrow}$ that it is possible to reproduce the modulation observed in the experiments. 
%

\end{document}